\documentstyle[11pt,newpasp,twoside,epsf]{article}

\begin{document}
\title{2D spectroscopy of double-barred galaxies}
\author{ A.V. Moiseev}
\affil{ Special Astrophysical Observatory, Nizhnij Arkhyz,
 Karachaevo-Cherkesia, 369167, Russia }
\author{J.R. Vald\'es, and V.H. Chavushyan}
\affil{Instituto Nacional de Astrof\'{\i}sica Optica y Electr\'onica, Apartado Postal 51 y 216, C.P.
72000, Pueblo, Pub., M\'exico}

\begin{abstract}
 The first results of the observational program of the study of
2D-kinematics in  double-barred galaxies are presented. We show that,  for the main part of the sample,
the inner bars
 do not affect the circumnuclear stellar kinematics.
Therefore, they are
 not dynamically decoupled structures.
Various types of non-circular gas motion were found in many galaxies.  The analysis of the ground-based
and HST optical and NIR images reveals mini-spirals in about half of the investigated objects. We
suggest that so called ``double-barred galaxies'' are, in fact, galaxies with very different
circumnuclear structure.
\end{abstract}

\section{Double-barred galaxies: history of the problem.}

Quarter of century ago, de Vaucouleurs (1975) found  that an inner  bar-like structure was nested in a
large-scale bar at the optical image of NGC\,1291. The first  systematic observational study of this
phenomenon was made by Buta \& Crocker (1993) who published a list of 13 galaxies, in which the
secondary bar was arbitrarily oriented in relation to the primary one. In the following years  the
isophotal analysis technique had allowed to  detect secondary bars at the optical (Wozniak et al.,
1995; Erwin \& Sparke, 1998) and at the NIR images of barred galaxies (Friedli et al., 1996; Laine et
al., 2001). Up to now, we have found, in total, 71 candidates to double-barred galaxies in the
literature (Moiseev, 2001). Also numerous theoretical papers exist, but secondary bar dynamics is still
far from being understood. Shlosman, Frank \& Begeman (1989) have shown that a new circumnuclear bar
may be  formed as a result of gas instabilities within a primary bar. Maciejewski \& Sparke (2000) found
some families of closed orbital loops which can support both bars. Dynamics of stellar-gaseous secondary
bars have been studied in numerical simulations (Pfenninger \& Norman, 1990; Friedli \& Martinet,
1993). New hydrodynamical simulations of the gas behaviour in double bars are presented by Heller,
Shlosman \& Englmaier (2001), Maciejewski et al. (2002) and Shlosman \& Heller (2001). To test these
theoretical predictions, new observational data is required.

 In this paper we try to find an answer to the question
 -- {\it Are the secondary bars dynamically decoupled systems?}

\begin{figure}[t]
\plotone{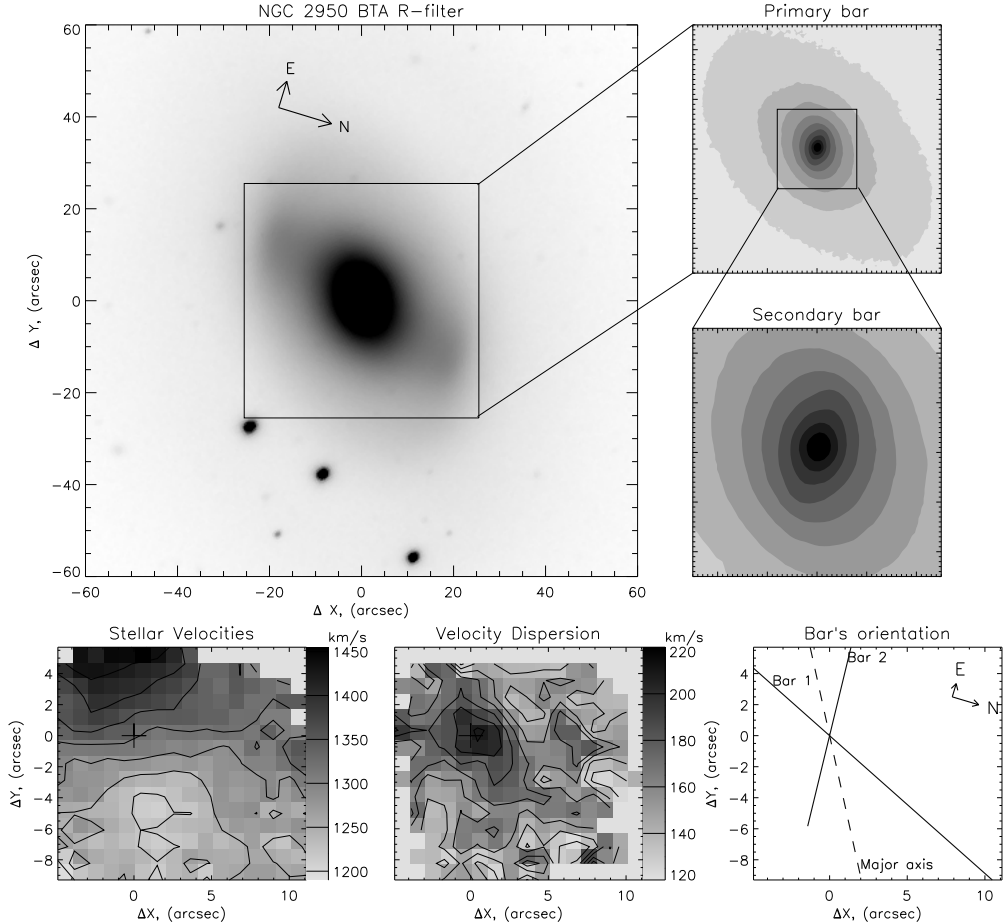} \caption{{\it top} -- the R-band image of NGC~2950.  {\it bottom} --
stellar velocity field, velocity dispersion distribution and sketch for  orientation of the bar. }
\label{n2950}
\end{figure}

\section{Necessity of 2D spectroscopy}
Mostly of the information about double-barred objects has been obtained from galaxy imaging (optical or
NIR). However the photometric structure revealed for them can be explained in a less exotic manner,
without involvement of the secondary bars. For example, the following explanations of the  turn of the
inner isophote major axis are possible:

\begin{itemize}
   \item a dynamically independent secondary bar ($\Omega_P\ne\Omega_S$,
   where $\Omega_P$ and
   $\Omega_S$ are angular rotation speeds of the primary and secondary bars respectively);
   \item dynamically coupled bars $(\Omega_P=\Omega_S)$, or
   $x_2$-orbits between two inner Lindblad resonances (ILRs) in the primary bar;
   \item  an elliptical ring in the plane  of the disk, at the ILR resonance
   of the primary bar;
   \item  a  polar disk (or an inner polar ring) within the primary bar
   (similar structures have been found in a number of ordinary early-type
   disk galaxies, see Afanasiev \& Sil'chenko, 2000; Sil'chenko, this
   proceeding);
   \item a the projection of the central part of an oblate bulge inside
   the primary bar onto the sky plane;
   \item  a complex distribution of dust and star formation regions
   in the circumnuclear region.
\end{itemize}

It seems to us that additional observations of the stellar and gas kinematics in these strange objects
might prove if the secondary bars are new dynamically isolated structures in barred galaxies.
Nevertheless, such observations are very rare (Wozniak, 1999; Emsellem et al., 2001). Emsellem et al.
(2001) have presented long-slit stellar kinematics of 4 double-barred objects. However two-dimensional
distributions of velocities and velocity dispersions can give us a more detailed information because
the objects observed are non-axisymmetrical.

 \section{ Observations}
In 2000 we observed a sample of the northern double-barred galaxies at the 6m telescope of the SAO RAS
by using various 2D spectroscopy methods. The circumnuclear regions were investigated with the
integral-field spectrograph MPFS (Afanasiev et al., this Proceeding). The velocity fields of the ionized
gas in the [OIII]$\lambda5007$ and $H_\beta$ emission lines, the stellar line-of-sight velocity and
velocity dispersion (hereafter $\sigma$) fields were constructed over $16\arcsec \times 15\arcsec$
field of view. For some galaxies we obtained large-scale ionized-gas velocity fields in the
[NII]$\lambda6583$ and/or $H_\alpha$ emission lines by using the scanning Fabry-Perot Interferometer
(IFP). The analysis of the morphology of the galaxies was done by means of JHK' images obtained with
CAMILA (Cruz-Gonzalez et al., 1994), at the 2.1m telescope (OAN, Mexico).  Optical images from the 6m
telescope and high-resolution images from  the HST Archive were also used for the same purpose.

\section{Results}

Fig.~\ref{n2950} shows our data for  NGC~2950. It is an SB0 galaxy, at the images of which the inner
bar-like structure is clearly seen that is turned by approximately $60^\circ$ with respect to the
primary bar. This ``secondary bar'' was firstly described by Wozniak et al. (1995) and Friedli et al.
(1996). Surprisingly, the position angle (hereafter PA) of the kinematical axis defined from the
circumnuclear stellar velocity field, $PA_{KIN}=125\pm 3^\circ$, is not significantly different from the
orientation of the outer isophotes of the disk, $PA_{DISK}=120\pm5^\circ$. This fact is
unexpected\footnote{Our present value of the $PA_{DISK}$ obtained from the deep R-band image disagrees
with the $PA_{DISK}=110^\circ$ from Friedli et al.(1996). The latter value was also used by Moiseev
(2001) that has resulted in a erroneous conclusion on the significant difference between $PA_{KIN}$ and
$PA_{DISK}$ in NGC~2950.} because the secondary bar-like structure is sharply distinguished at the
galactic images and must affect the stellar orbital motions if it is triaxial.

The central ellipsoidal distribution ($r\sim 5\arcsec$) is seen at the velocity
dispersion map (see Fig.~\ref{n2950}), with the major axis of this distribution being
coincident with the outer (not with the inner!) bar orientation. This situation is
typical for single-barred galaxies, where distributions of line-of-sight $\sigma_{*}$ in
the  bar would be symmetric to the bar axis but not to the major axis of the disk
(Miller \& Smith, 1979; Vauterin \& Dejonghe, 1997). In  other objects observed we have
found more complex velocity dispersion distributions than regular elliptical peaks (see
Moiseev, 2001). These are peaks shifting from the dynamical centers or central velocity
dispersion minima. The latter may be connected with a presence of a cold dynamical
component (Emsellem, 2001) or with the evolution of the stellar bars (see simulations by
Khoperskov, Moiseev \& Chulanova, 2002).

\begin{figure}
\plotone{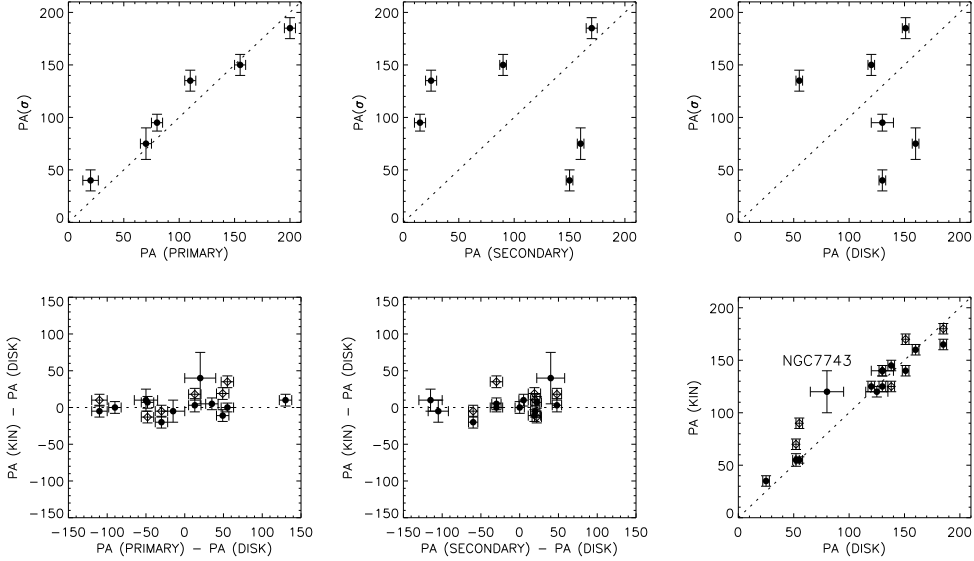} \caption{Position angles of the kinematical features versus PAs of the
main directions in the observed galaxies. Stellar and ionized gas data are plotted by black and open
symbols. Dotted lines mark the direct proportionality of PAs (not the best fit!)} \label{stat}
\end{figure}

We have considered some statistical properties of the our sample. If a secondary bar is truly decoupled
then the galaxy has three {\it main directions}: $PA_{DISK}$ of a disk major axis, and also
$PA_{PRIMARY}$ and $PA_{SECONDARY}$ that are traced by both bars in the sky plane. Fig.~\ref{stat}
shows the $PA$ of the observed {\it kinematical features} versus the $PA$ of these main directions.
Here the kinematical features are $PA_{\sigma_{*}}$ -- the orientation of the elliptical
$\sigma_{*}$-peak (if it is present) and  $PA_{KIN}$ -- orientation of the central stellar kinematics
axis. One can clearly see that the stellar velocity dispersion distributions correlate only with the
{\it primary} bars.  The secondary bars have not any influence on the stellar kinematics.

 The bottom plot in Fig.~\ref{stat} reveals two important facts. Firstly,
 the deviations of $PA_{KIN}$ and $PAs$ of the both bars from the galactic
 major axis do not correlate. Secondly, one can see
 that the circumnuclear kinematical axes, deduced from the velocity
fields, correlate with the global disk orientations. Therefore both gas and stars
orbital motions are mostly axisymmetrical in the nuclei of the observed galaxies. The
residual deviations $PA(KIN)$ from $PA_{DISK}$ could be connected with non-circular gas
motions. A detailed consideration of the ionized gas kinematics in individual objects
reveals  various types of such motions. For example, the large-scale (primary) bar in
NGC~470 provokes non-circular gas flows in the IFP velocity field. At the secondary bar
distance ($r<8\arcsec$) the isovelocity turn may be explained by a radial gas inflow in
the bar or by a polar disk appearance within the primary bar. The comparison with the
isophotal analysis results allows us to choose the disk instead of the bar.

\begin{table}
\caption{Circumnuclear properties of  double-barred galaxies} \label{tab1}
\begin{tabular}{lrlll}
\tableline
        &  & \multicolumn{2}{c}{Kinematics}  & NIR/opt. morphology \\
Name   &   & distribution of $\sigma_{*}$ & ionized gas &    \\
 \tableline                                                         \\
NGC&   470   &ellipse &polar disk& --    \\
NGC&  2273   &ellipse  &flows in P-bar&  ring and pseudo spiral  \\
NGC&  2681   & ellipse & -- & -- \\
NGC&  2950 & ellipse & --&--\\
NGC&  3368  & non-central peak & bar & irregular dust spiral\\
NGC&  3786 & two peaks  & flows in P-bar&--  \\
NGC&  3945  &ellipse &--& embedded  ring\\
NGC&  4736   & central drop &  flows in P-bar& flocculent spiral     \\
NGC&  5566   & -- & -- & two-armed spiral    \\
NGC&  5850    & --  & polar disk& embedded disk\\
NGC&  5905  & ellipse & inflow in bar&  --  \\
NGC&  6951  & --& radial motions&  ring and flocculent spiral\\
NGC&  7743    & --& -- & flocculent spiral \\
\tableline
\end{tabular}
\end{table}

In the table~\ref{tab1} we summarize all the circumnuclear properties that were found in our sample.
Although various kinematical features are present, only in the case of NGC~3368 the gas kinematics may
be explained by a dynamically decoupled secondary bar.

\begin{figure}
 \plotone{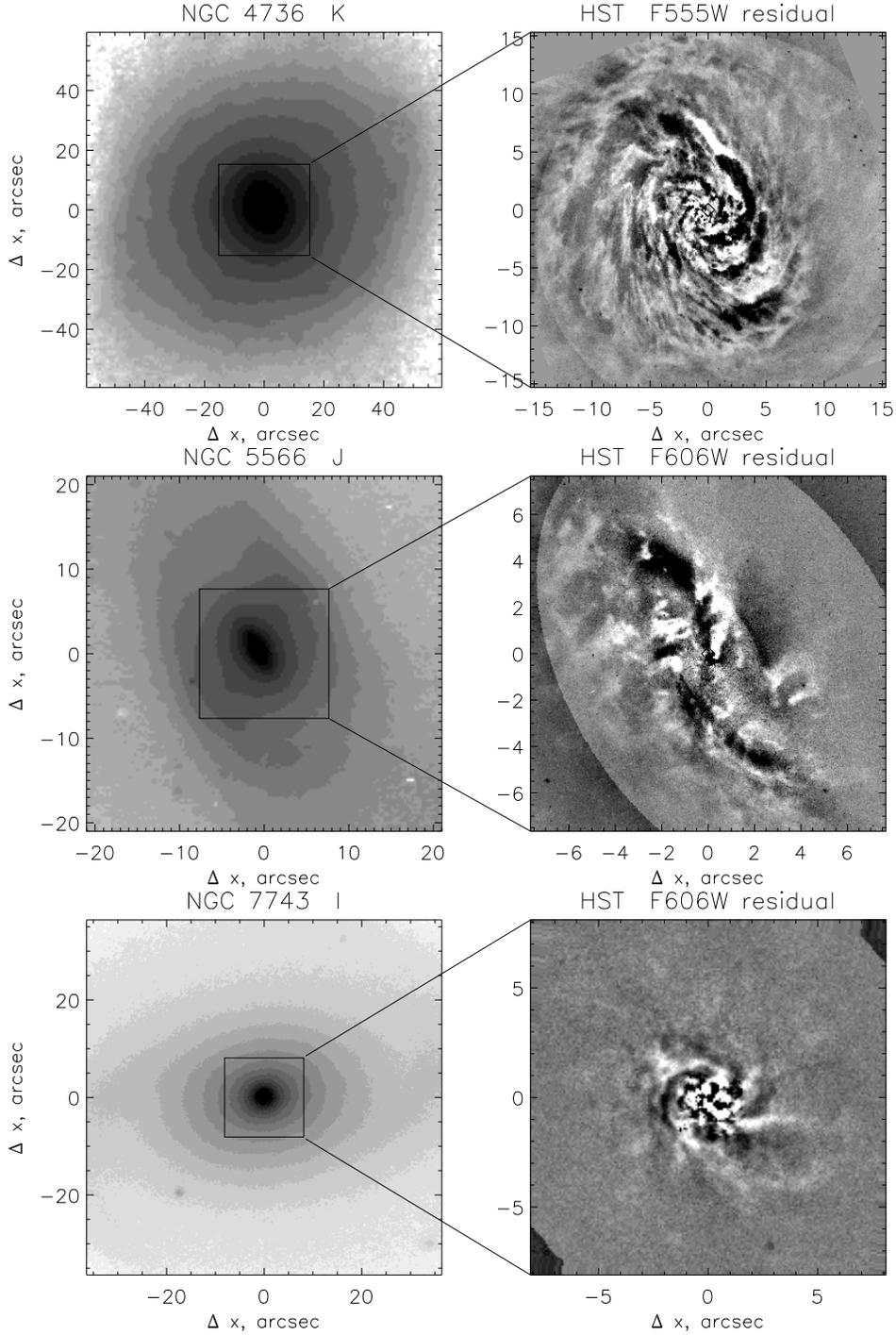} \caption{{\it left} -- NIR images of  NGC~4736 and
NGC~5566 and I-band image of NGC~7743. {\it right} --  the HST Planetary Camera  images, for the same
objects,  after elliptical isophotes removal.} \label{spiral}
\end{figure}

\section{Do mini-spirals nest in the primary bars?}

In order to study the residual brightness morphology of the galaxies, we have subtracted elliptical
isophotal models from the optical and NIR images. Interestingly, about half of all galaxies have
mini-spirals or pseudo-spirals around their nuclei. Three examples of  nuclear spirals are shown in
Fig.~\ref{spiral}. The multi-armed flocculent spiral in  NGC~4736 is studied also  by Elmegreen,
Elmegreen \& Eberwein (2002). The two-armed spiral in NGC~5566 resembles model calculations of the gas
behaviour in a stellar bar  by Englmaier \& Shlosman (2000). P\'erez et al. (2000) found a flocculent
mini-spiral structure on the HST sharp-divided images inside $r<6\arcsec$ in NGC~6951. We suggest that
this spiral causes a unusual residual (non-circular) velocity picture that was observed at our
Fabry-Perot velocity field of NGC~6951.

In the recent years, the interest for the nuclear spirals has considerably grown, because these
features have been found in the high-resolution HST images of some galaxies(Van Dokkum \& Franx (1995),
Elmegreen et al.(2002) and references there). The origin of these structures is yet unknown, different
models are proposed in the literature. In our sample a  clear connection is seen between the
large-scale bars and mini-spiral structure within them.

\section{ Conclusions}

A sample of double-barred galaxies has been observed at the 6m telescope by the 2D-spectroscopy methods.
In order to know about the morphology of the sample galaxies, we also carried out NIR imaging
observation with CAMILA, at the 2.1m telescope (OAN, M\'exico) The velocity fields of the ionized gas
and stars and the line-of-sight stellar velocity dispersion maps are constructed and analyzed. The
following  features are found:

\begin{itemize}
\item  The line-of-sight velocity dispersion distributions of stars
are connected with the primary bars only. Also the circumnuclear stellar kinematical axes do not show
significant misalignment with the $PA$'s of the main discs. Therefore the secondary bars are not
dynamically decoupled stellar systems.

\item  Remarkable non-circular motions of the ionized gas are found
in $\sim60\%$ of all objects. These features may be associated with
 radial gas inflow in the  primary bar, or with non-circular motions
 between ILRs of the  primary bar, or with polar (inclined) circumnuclear
 discs, or with gaseous motions in the circumnuclear mini-spirals.

\item The analysis of the optical and NIR images shows that mini-spiral
structures exist at the secondary bar radial scales in 6 out of 13 objects.

\end{itemize}

\noindent Thus we suggest that a dynamically independent secondary bar is a much more rare phenomenon
than it  follows from the isophotal analysis of images itself.

\acknowledgements{We~~~ would~ like~~ to~~ thank~~ V.L. Afanasiev and  O.K. Sil'chenko
for their interest and numerous discussions. This work was partially supported by the
RFBR grant (Project No. 01-02-17597) and by the Federal program "Astronomy" (Project
1.2.3.1). This work is based on the observations from the 6m telescope of the Special
Astrophysical Observatory of the Russian Academy of Sciences operated under the
financial support of the Science Department of Russia (registration number 01-43). The
authors would like to thank all the staff of the {\it Observatorio Astron\'omico
Nacional}, in San Pedro Martir, M\'exico for their assistance during the NIR
observations}

\end{document}